# Dynamic Pattern Based Image Steganography

P.Thiyagarajan, G.Aghila, V.Prasanna Venkatesan

**Abstract**—Steganography is the art of hiding secret information in media such as image, audio and video. The purpose of steganography is to conceal the existence of the secret information in any given medium. This work aims at strengthening the security in steganography algorithm by generating dynamic pattern in selection of indicator sequence. In addition to this dynamicity is also encompassed in number of bits embedded in data channel. This technique has been implemented and the results have been compared and evaluated with existing similar techniques.

——————————— ◆ ———————————

## 1 INTRODUCTION

Information is the wealth of any organization. This makes security a major concern in today's world. Cryptography is a technique for securing the secrecy of communication and many methods have been developed to encrypt and decrypt the messages. Unfortunately it is sometime not enough to keep the contents of message secret it may be also necessary to keep the existence of message secret which is done by steganography. Steganography involves hiding information inside any cover media (image or audio or video file[5]) such that it appears no message is hidden[9]. Several methods have been proposed for image based steganography such as LSB [4], RGB single and multi channel embedding.

In this paper we propose a Dynamic Pattern based Image Steganography (DPIS) technique for RGB based image steganography. DPIS technique ensures minimum capacity for storing secret message than existing techniques and it also ensures that channels containing relatively lower colour values can store more bits of data. Experimental results show that our technique performs much better than the existing techniques. The paper is organized as follows Section 2 deals with brief description of existing techniques, Section 3 explains the proposed technique Section 4 depicts the evaluation criteria and experimental results and Section 5 concludes the paper.

## 2 RELATED WORK

There are many methods exists for image steganography [8] such as Fibonacci data hiding technique, DCT Algorithm ,Data hiding using Prime numbers , etc. This section focuses on the algorithms which uses RGB channel for Steganography.

Adnan Gutub et.al [1] uses any one of the channels as Indicator channel. Indicator channel indicates the data channel among the other two channels. The Indicator channel is chosen in sequence with 'R' being the first.

- If the last two bits of the indicator are '00' then no data is embedded in the channel 1 and channel 2.

- If the last two bits of the indicator are '01' the no data is embedded in channel 1 and 2 bits of data is embedded in channel 2.

- If the last two bits of the indicator are '10' then two bits of data is embedded in channel 1 and no data is embedded in channel 2.

- If the last two bits of the indicator are '11' then two bits of data is embedded in channel 1 and channel 2.

————————————————

*P.Thiyagarajan is with the CDBR- SSE Lab Department of Computer Science, Pondicherry University, R.V.Nagar, Kalapet Puducherry – 605 014.   E-mail: thiyagu.phd@gmail.com*
*G.Aghila is with the CDBR- SSE Lab Department of Computer Science, Pondicherry University, R.V.Nagar, Kalapet Puducherry – 605 014.   E-mail: aghilaa@gmail.com*
*V.Prasanna Venkatesan is with the CDBR- SSE Lab Department of Computer Science , Pondicherry University, R.V.Nagar, Kalapet Puducherry – 605 014.   E-mail: prasanna_v@yahoo.com*

Mohammad Tanvir Parevez et.al [7] uses the concept lower color value of a channel has less effect on the color of a pixel than the higher value therefore more data bits can be embedded in the lower color channel. The Number of bits embedded in each channel depends on partition schema which is static through out the embedding process. Among the R, G and B channel any channel is chosen as Indicator channel and it can be made random. Indicator channel sequence rotates in circular way through the embedding process.

The above listed techniques suffers from the following limitations

- Data are embedded sequentially in all pixels

- Data are embedded using static partition schema or same number of bits in the channel

- Fails to detect whether the stego image has been modified by the intruder.

DPIS technique overcomes these limitations and it has been proven by experimental results.

## 3 Dynamic Pattern Based Image Steganography (DPIS) Technique

In this section Dynamic Pattern based Image Steganography technique (DPIS) was discussed in detail. The idea behind our technique is that dominant color or Significant color in a pixel should not suffer from data embedding while the insignificant color channel can be used for data embedding. Figure 1 and Figure 2 depicts the embedding and extracting process of DPIS.

Major steps in the DPIS technique

- Generate the random string of length say 'N' in terms of R, G and B.

- From the generated indicator sequence, choose the indicator channel and check whether the indicator channel is the lowest among the other channels. If Indicator is lowest, then skip the pixel from embedding else embed the data.

- Choose the data channel from the remaining two channel

- The number of bits to be embedded in data channel is determined at run time and it is not pre determined.

At the extraction part, channel in which data is embedded is commuinicated in the LSB of Indicator channel.

- The indicator sequence should be obtained from the embedding part.

- If the LSB of Indicator channel is '0', the channel follows immediately after the indicator is the data channel

- If the LSB of Indicator channel is '1' , the channel precedes the  indicator is the data channel

To extract the exact number of bits embedded in data channel we need to know

- If the LSB of third channel is '0', then three bits of data are embedded in data channel

- If the LSB of third channel is '1' , the four bits of data are embedded in data channel

DPIS technique has been tested on different image category such as portrait, flower, nature, toys etc. Figure 3 to Figure 5 shows the Cover and Stego images generated through DPIS technique.

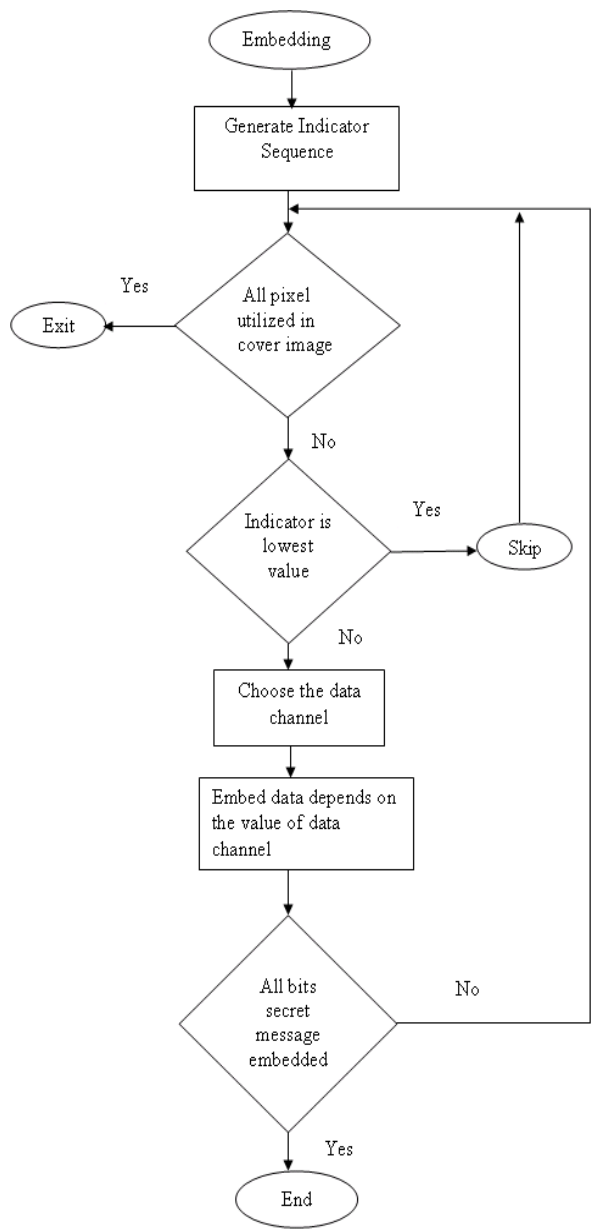

Fig. 1. Flow chart for embedding

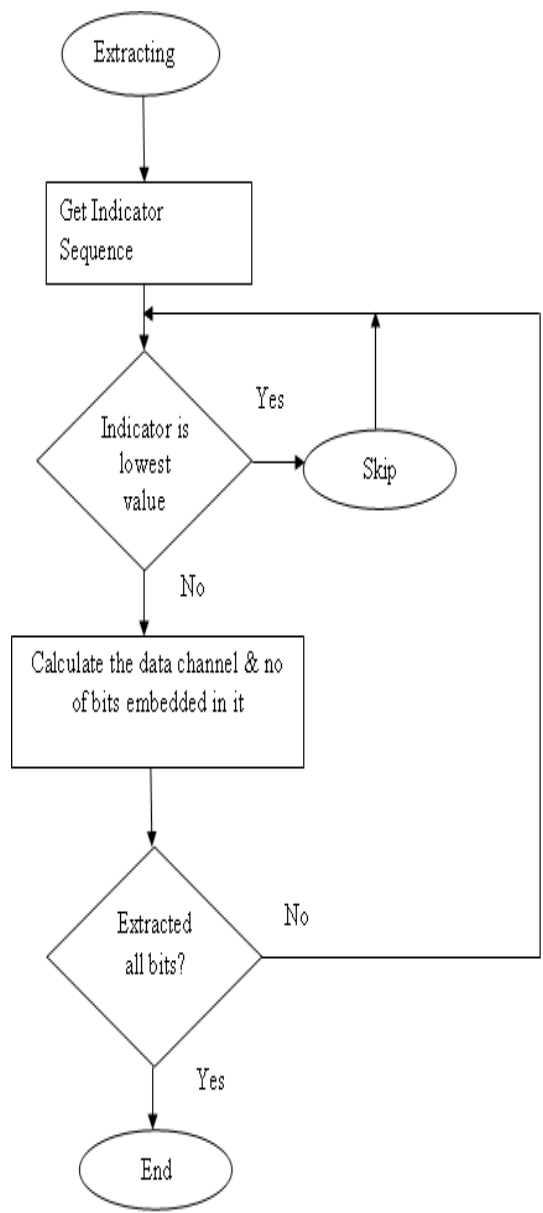

Fig. 2. Flow chart for extracting

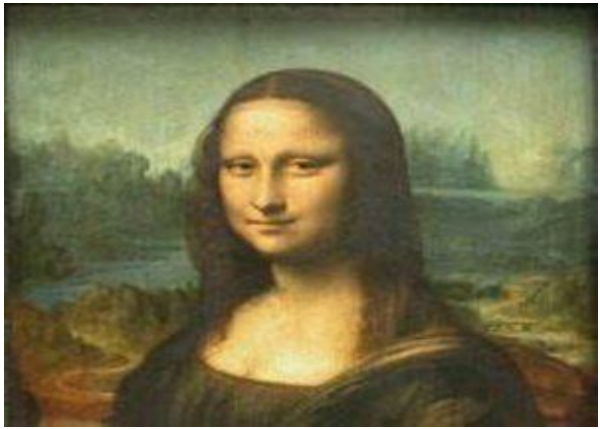

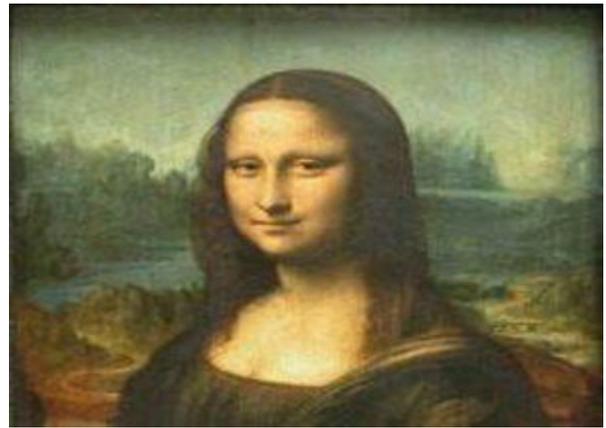

Fig. 3 (a). Cover Image, Image Size: 300 x 266

Fig. 3 (b). Stego Image, Pixel used for embedding: 3486, Secret Message Size: 15358

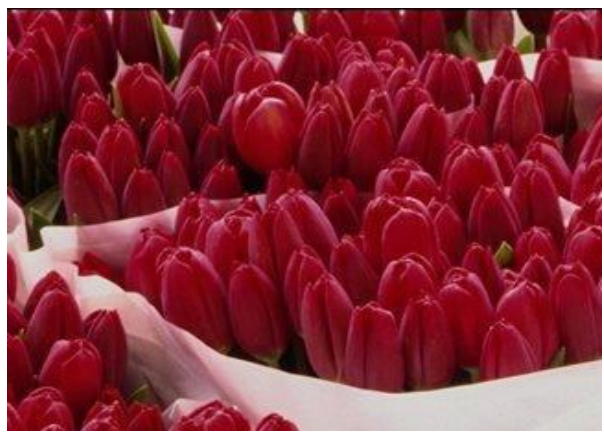

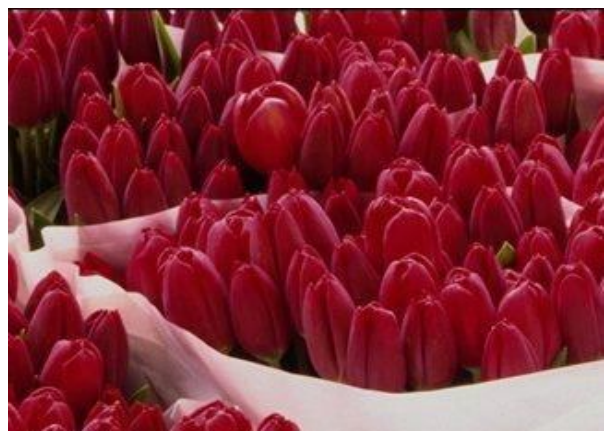

Fig. 4 (a). Cover Image, Image Size: 323 x 429

Fig. 4 (b). Stego Image Pixel used for embedding: 4714, Secret Message size: 21133

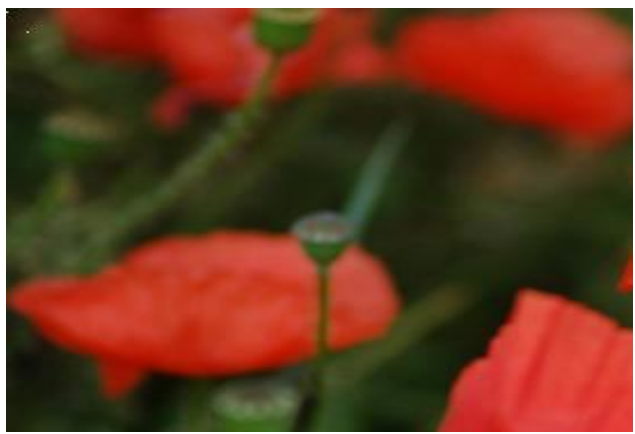

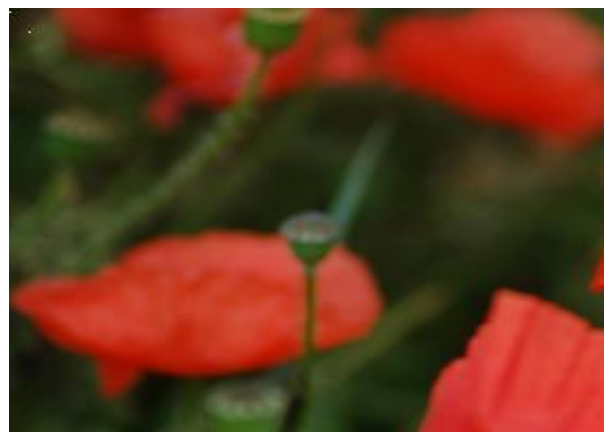

Fig. 5 (a). Cover Image, Image Size: 274 x 255

Fig. 5 (b). Stego Image Pixel used for embedding: 6298, Secret Message Size: 27755

# 4 EVLAUTION CRITERIA AND EXPERIMENTAL RESULTS FOR PROPOSED TECHNIQUE

DPIS technique is implemented in MATLAB and evaluated against the existing [1] [5] [7] and proposed parameters.

a) Robustness
b) Capacity
c) Invisibility
d) Behaviour of proposed technique on common steganalysis methods
e) Detection of tampered stego images

## 4.1 Robustness

Robustness is measured by the difficulty level of the intruder to break the key for any technique. Given any RGB image intruder may try for brute force attack for locating the Indicator. If the indicator is identified then data channel can be easily traced out and secret message can be extracted or cramped.

The number of distinct pattern of indicator sequence shown in Table 1 is obtained by the formulae $3^{(n-2)}*2$ where n is the length of the indicator sequence.

For all our experiments the indicator length of 20 have been choosen that generates 7, 74,840,978 distinct indicator sequences, which is very difficult to break by brute force attack.

TABLE 1
LENGTH OF INDICATOR SEQUENCE AND NUMBER OF DISTINCT PATTERNS

| Length of the Indicator | Number of distinct pattern |
|---|---|
| 3 | 6 |
| 10 | 13,122 |
| 15 | 3,18,8646 |
| 20 | 7,748,40,978 |

## 4.2 Capacity

Capacity is key important evaluating parameter in steganography technique. Capacity is defined as number of pixels used in the cover image to embed the secret message of any length. Our DPIS technique uses very minimum number of pixels compared to the existing technique and this has been proved by experimental results in Table 2.Graphical representation of the table 2 is shown in the figure 6.

TABLE 2
COMPARISON OF PIXEL UTILIZATION BETWEEN EXISTING AND PROPOSED TECHNIQUE

| Techniques | Size of secret message in bits | Number of pixels in Cover image | No of utilized pixels for embedding in cover image | % of used pixels in cover image |
|---|---|---|---|---|
| RGB Intensity based variable bits | 15,358 | 79,800 | 7154 | 8.96 % |
|  | 21,133 | 69,870 | 11485 | 16.43% |
|  | 27,755 | 80,429 | 15981 | 19.86% |
| Pixel Indicator | 15,358 | 79,800 | 7500 | 9.39% |
|  | 21,133 | 69,870 | 12,099 | 17.31% |
|  | 27,755 | 80,429 | 16,617 | 20.66% |
| DPIS | 15,358 | 79,800 | 3486 | 4.36 % |
|  | 21,133 | 69,870 | 4714 | 6.74% |
|  | 27,755 | 80,429 | 6298 | 7.83 % |

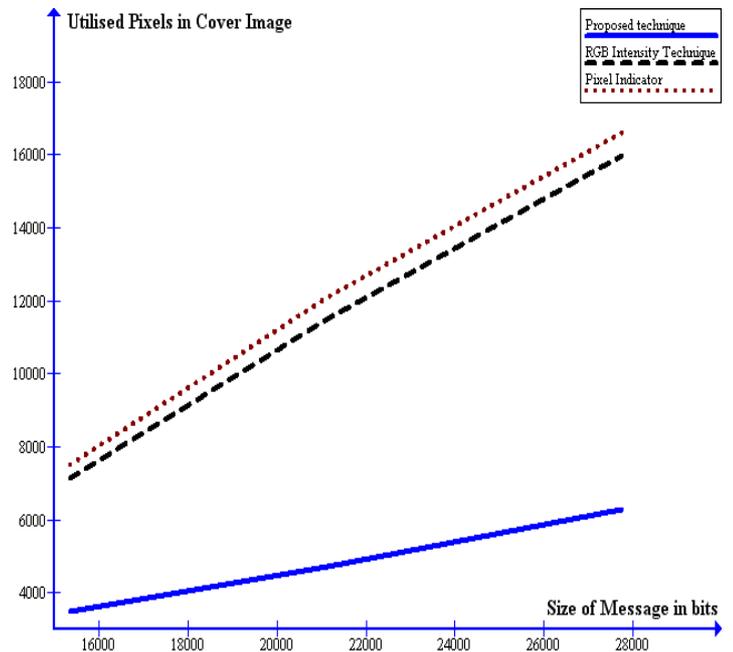

Fig. 6. Graph showing pixel utilisation by existing and proposed technique

## 4.3 Visibility

Once the message has been embedded in cover image changes in the pixel value should not be noticed in the stegoimage. It is very hard to find the small color change in the cover and stego image by human eye. For this histogram comparison is used by steganalyst to identify the stego-image by comparing the histogram of cover image and stegoimage.

Based on this experiments have been conducted and separate histograms are drawn for cover and stego image .It is observed that there is no major difference visible while comparing Red, Green and Blue histogram of both cover and Stego Images. Figure 7 -8 shows the histogram of RGB channels.

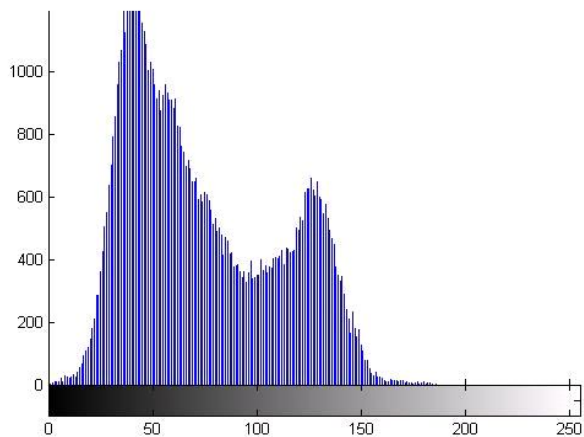

Fig. 7 (c). Original Image Blue Histogram

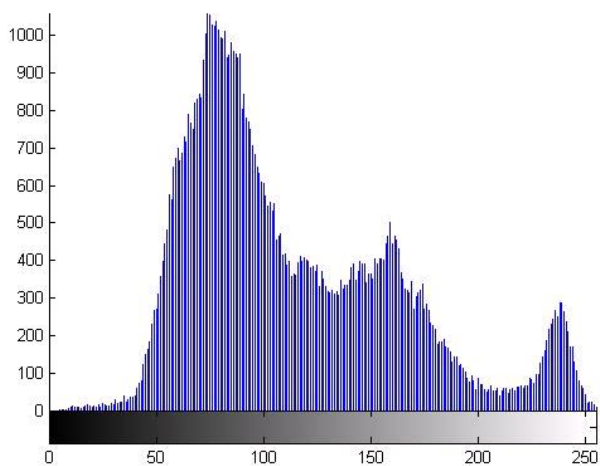

Fig. 7 (a). Original Image Red Histogram

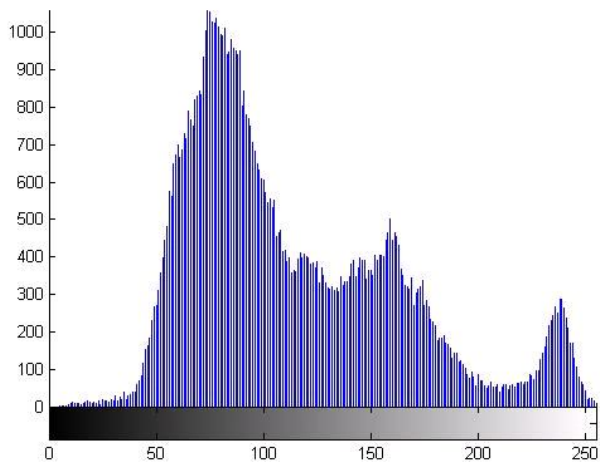

Fig. 8 (a). Stego Image Red Histogram

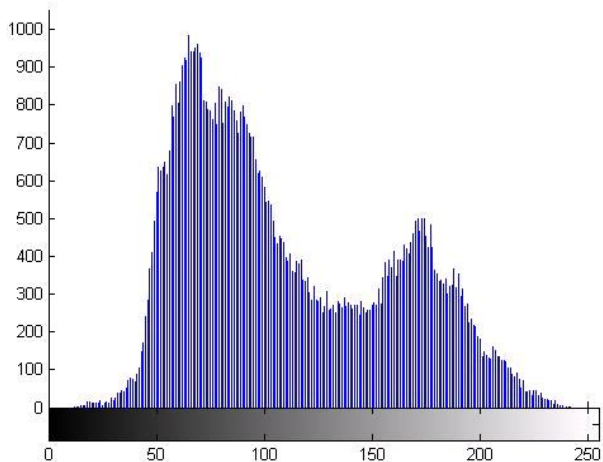

Fig. 7 (b). Original Image Green Histogram

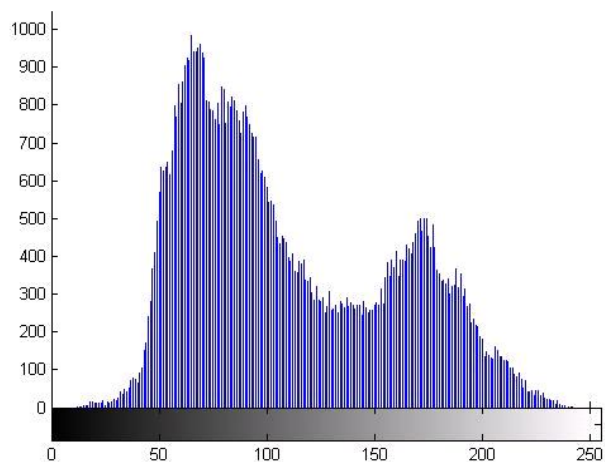

Fig. 8 (b). Stego Image Green Histogram

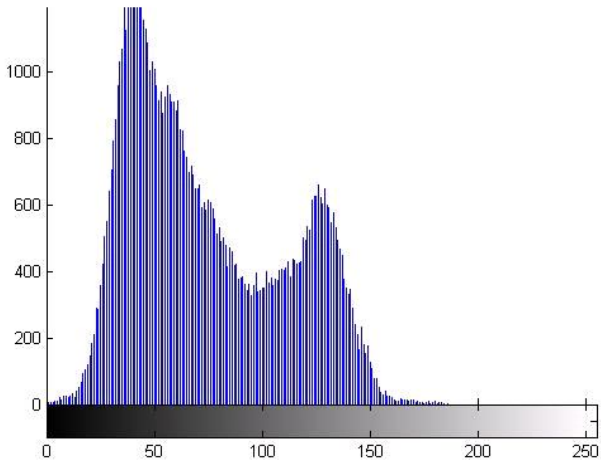

Fig. 8 (c). Stego Image Blue Histogram

### 4.4 Behavior of proposed technique on common steganalysis methods

Steganalysis is the art and science of detecting messages hidden using steganography [10]. The most common steganalysis methods used for RGB based steganography are

- Brute force attack on Indicator sequence
- Extracting data from all pixels
- Extracting same number of bits from data channels

Brute force attack on Indicator sequence:

Brute force attacks for steganography [3] involves in trying all possible keys until valid key is found. In dynamic pattern based image steganography technique, indicator channel contains the information where the data are stored. Intruder may try the brute force attack on indicator sequence until meaningful message is traced. In all our experiments length of indicator sequence is greater or equal to 20 so number of distinct pattern generated is very high and it cannot be broken by brute force attack.

The secret message shown in the Table 3 has been embedded in the cover medium by DPIS technique and stegoimage obtained is shown Figure 9. Brute force attack has been applied on the indicator sequence for extracting message from the stego image.

TABLE 3
EXPERIMENT RESULT : EMBEDDED MESSAGE AND EXTRACTED MESSAGE WITH WRONG PIXEL INDICATOR

| Embedded Secret Message in cover medium | Secret message obtained by wrong indicator sequence |
|---|---|
| PondicherryUniversity Computer Science Dept | QibP(97u2q□□<□<]□4]nD \|:8*□□~v&N-F□KKe□W;' |

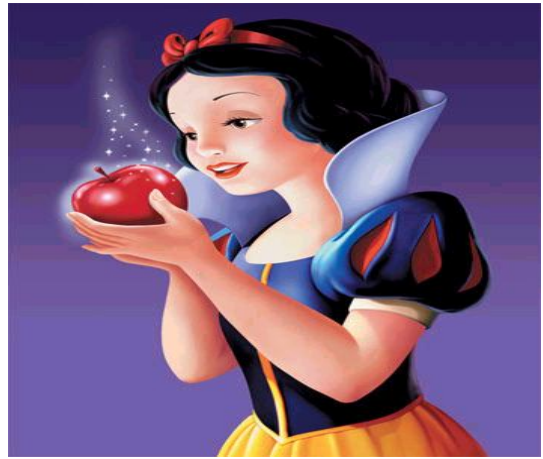

Fig 9: Stegoimage generated by DPIS algorithm

The above experiment depicts that even if one value in the indicator sequence is not correct, embedded secret message cannot be extracted.

Extracting data from all the pixels sequentially:

In DPIS method data are not embedded in all the pixels sequentially. Some pixels in the sequence are missed inorder to strengthen the technique. The Table 4 shows the result of extracting data from all the pixels from stegoimage shown in Figure 10.

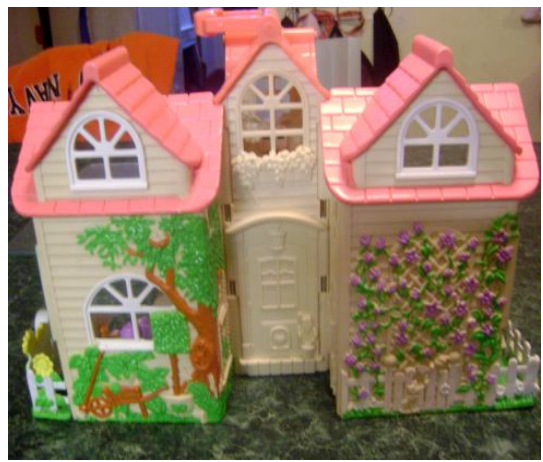

Fig 10: Stegoimage generated by DPIS algorithm

TABLE 4
EXPERIMENT RESULT : EMBEDDED MESSAGE AND EXTRACTED MESSAGE FROM ALL THE PIXELS IN THE STEGO IMAGE

| Embedded Secret Message in cover medium | Secret message obtained by extracting bits from all pixels in stego image |
|---|---|
| Pondicherry University Computer Science Dept | T+*m*☐-&#tKK`w0    -xy |

Extracting same number of bits from all pixels:

In the DPIS technique the number of bits embedded in each pixel varies and it is decided on the run time. Experiments were conducted for extracting same number of bits from all the pixels in the stegoimage generated by DPIS technique. The Table 5 shows the result of extracting same number of bits from all the pixels from stegoimage shown in Figure 11.

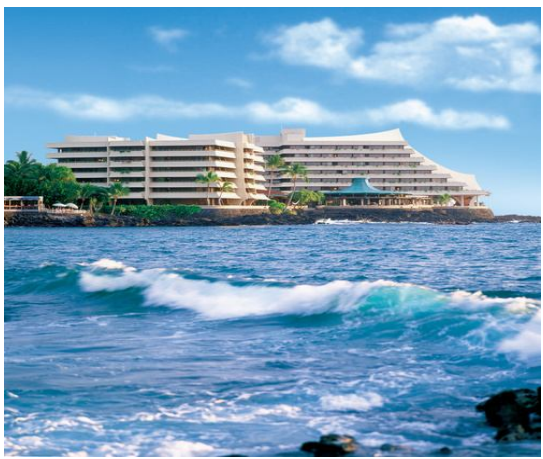

Fig 11: Stegoimage generated by DPIS algorithm

TABLE 5
EXPERIMENT RESULT : EMBEDDED MESSAGE AND MESSAGE EXTRACTED WITH UNIFORM NUMBER OF BITS FROM PIXELS IN THE STEGO IMAGE

| Embedded Secret Message in cover medium | Secret message from stego image obtained by extracting 2 bits from all data channels |
|---|---|
| Pondicherry University Computer Science Dept | Sj Lc;G?Cq chT=Y{ w@6O` =o `4'h \|    -xy |

## 4.5 Detection of tampered stego images

In DPIS technique, provision has been given to check whether the stego image has been modified by any intruder [2]. Before decrypting the embedded message the stegoimage is checked for any change done by intruder. If stegoimage is not tampered then the extraction part will be executed.

This has been achieved in the DPIS technique by communicating the information about unmodified pixels and their values to the counter part. Table 6 shows the comparative study of some important features between DPIS and existing techniques.

TABLE 6
EXPERIMENT RESULT : COMPARATIVE STUDY OF SOME IMPORTANT FEATURES BETWEEN PROPOSED AND EXISITING TECHNIQUES

| Features | RGB Intensity based variable bits | Pixel Indicator | DPIS technique |
|---|---|---|---|
| Indicator sequence | Vary in length of 3 | Static | Varying with different length |
| Brute force attack | Breakable | Breakable | Not Breakable |
| Number of bits embedded in data channel | Static and depends on partition schema | Static | Dynamic and its decided at run time |
| Sequential data embedding in all pixels | Yes | No | No |
| Tampered stego image detection at decryption part | Not possible | Not possible | Possible |

## 4 Conclusion

In this paper, dynamic pattern based image steganography technique has been introduced. This technique addresses key important issues like dynamicity in data embedding and indicator sequence and thus making it difficult to hack by steganalyst. Moreover it also detects whether the stego image has been modified by intruder during transmission. DPIS technique results in very high capacity with low visual distortions and all this have been proved by experimental results. DPIS technique has also been compared with important feature of other steganography algorithms

## Acknowledgment

This research, CDBR Smart and Secure Environment, was sponsored by National Technical Research Organisation (NTRO) and their support is greatly acknowledged

**Mr.P.Thiyagarajan** is currently working as Research Associate in CDBR- Smart and Secure Environment Research Lab and pursuing his Ph.D in Pondicherry University. He obtained his Integrated M.Sc Computer Science with distinction from College of Engineering Guindy Anna University Main Campus Chennai. Prior to joining research he has couple of years experience in Sotware Industry. His research interest includes Information security, Network Security and Database Management System.

**Dr.G.Aghila** is currently working as Professor in the Department of Computer Science, Pondicherry University. She obtained her undergraduate degree (B.E (CSE)) from TCE, Madurai in the year 1988. Her postgraduate degree M.E. (CSE) is from the School of Computer Science, Anna University, Chennai in the year 1991. She obtained her Doctorate in the field of Computer Science from the Department of Computer Science, Anna University, Chennai in the year 2004. She has got more than two decades of Teaching Experience and published 29 research publications in the National/ International journals and conferences. Her research interest includes Knowledge Representation and    reasoning systems, Semantic web, Information Security and Cheminformatics.

**Dr.V.Prasanna Venkatesan** is currently working as Associate Professor in Banking Technology Department under the School of Management at the Pondicherry University. Prior to joining the Department of Banking Technology, he was the Lecturer of Computer Science at Ramanujan School of Mathematics and Computer Science. He has published one book and more than thirty research papers in various journals, edited book volumes and conferences. He has earned B.Sc. (Physics) from Madras University and MCA, M.Tech (CSE) and PhD (CSE) from Pondicherry University, Pondicherry, India. His research interest includes Software Engineering, Object Oriented Modeling and Design, Multilingual Software Development, and Banking Technology.